\documentclass[aps,a4paper,showpacs]{revtex4}

\usepackage{psfrag,graphicx,color,subfigure}
\usepackage{enumerate}

\begin{document}

\title{Elasto-plastic flow of a foam around an obstacle}

\author{F. Boulogne}
\affiliation{Institute of Mathematics and Physics, Aberystwyth University, SY23 3BZ, UK}
\affiliation{Univ Paris-Sud, Univ Pierre et Marie Curie-Paris 6, CNRS,\\ Lab FAST, Bat 502, Campus Univ -- F-91405, Orsay, France.}
\author{S.J. Cox}
\email[E-mail: ]{foams@aber.ac.uk}
\affiliation{Institute of Mathematics and Physics, Aberystwyth University, SY23 3BZ, UK}

\date{\today}

\begin{abstract}
We simulate quasistatic flows of an ideal two-dimensional monodisperse foam around different obstacles, both symmetric and asymmetric, in a channel. We record both pressure and network contributions to the drag and lift forces, and study them as a function of obstacle geometry. We show that the drag force increases linearly with the cross section of an obstacles. The lift on an asymmetric aerofoil-like shape is negative and increases with its arc length, mainly due to the pressure contribution.
\end{abstract}


\maketitle

\section{Introduction}

Foams are used widely, for example in industries associated with mining, oil recovery and personal care products \cite{prudk96}. Their use is often preferred because of  properties such as a high surface area, low density and a yield stress \cite{WeaireH99,hohlerca05}. In addition to this evidence of plasticity, a foam's rheology is dominated by elasticity at low strains and viscous flow at high strain-rates: they are elasto-visco-plastic fluids \cite{mousse10}.

A common probe of foam rheology is a variation of Stokes' experiment \cite{stokes50} in which an object moves relative to a foam  \cite{coxahw00,asipausaggj03,bruyn04,dolletag04,dolleteqrag05,cantatp05,dolleteqhag05,coxdg05, cantatp06,dolletdg06,dolletg06,raufastedcjg06,tabuteauobc07,wyndc08,daviesc08}. Foams have an advantage over many complex fluids in that their local structure (the bubbles) is observable, thus making them an excellent choice to determine the mechanisms by which non-Newtonian fluids show different responses to Newtonian fluids. In addition, a two-dimensional foam is a realizable entity, for example the Bragg bubble raft \cite{braggn47}, with which it is possible to perform a rheological experiment in which the shape and velocity of each bubble can be tracked in time. Foams are also amenable to numerical simulation because of the precise local geometry that is found wherever soap films meet. Plateau's laws, which describe how the films meet, are a consequence of each soap film minimizing its energy, equivalent to surface area, and it is this that provides the algorithm for the work described here.

For flow to occur in a foam, the bubbles must slide past each other. This occurs through T1 neighbour switching topological changes \cite{weairer84}, in which small faces and/or short films disappear and new ones appear. Sometimes referred to as plastic events, these are a visible indication of plasticity in a foam, and act to reduce the stress and energy. Numerous contributions to viscous dissipation occur \cite{buzzlc95}, although we assume that if the flow is slow enough they can all be neglected.

%Amongst other well-known non-Newtonian effects such as rod-climbing \cite{tanner00}, 
\citet{dolletdg06} measured the drag, lift and torque on an ellipse in a two-dimensional foam flow in a channel. The lift was maximized when the ellipse was oriented at an angle of $\pi/4$ to the direction of flow.
\citet{dolletag04} found that an aerofoil embedded in a foam flow exhibited a {\em negative} lift, which they attributed to the elasticity of the foam. This augments the list of well-known non-Newtonian effects that contradict the sense of what is known for Newtonian fluids.

We present here elasto-plastic simulations, in the so-called quasi-static limit, for 2D foam flow around an obstacle, and investigate the effect of the symmetry of the obstacle in determining the magnitude and direction of the drag and lift. Such simulations allow us to exclude consideration of viscous effects, and even to separate out pressure and film network contributions to the forces on an obstacle, both of which are difficult to do in experiment. As a means of determining drag and lift on an obstacle, they have been validated against experiments on an ellipse \cite{dolletdg06} by \citet{daviesc10}.

\begin{figure}
\centerline{
\includegraphics{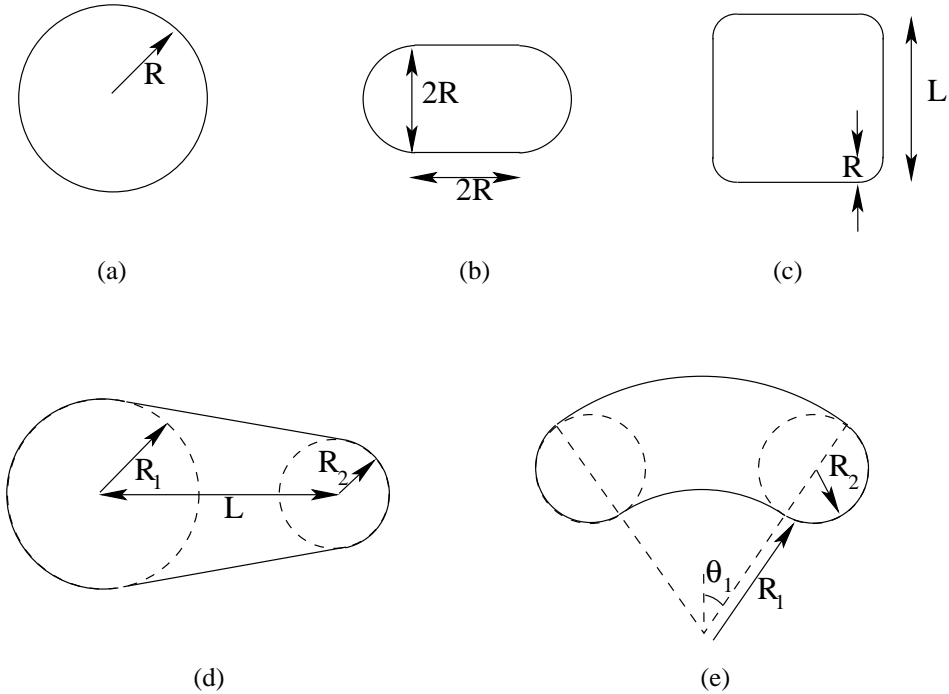}
}
\caption{Pictures of the obstacles, oriented with flow from left to right: (a) circle, (b) horizontal stadium, (c) square, (d) symmetric aerofoil and (e) asymmetric aerofoil.}
\label{fig:obstacles}
\end{figure}

We consider a range of obstacle shapes, illustrated in figure \ref{fig:obstacles}. Since the Evolver uses a gradient descent method, we are unable to simulate an obstacle with sharp corners. We therefore round the corners of each obstacle with segments of a circle to smooth the boundary. The shapes are: 
\begin{enumerate}[(a)]

\item a circle, which provides the standard case with full symmetry. Its cross-section is $H=2R$.

%\item an ellipse with eccentricity $e = 1.25$ (long axis parallel to the direction of foam flow) or $e=0.8$ (long axis perpendicular to the direction of foam flow) and $H=R/e$. 

\item the union of a square and two semi-circles, which we call a ``stadium'', arranged either vertically or horizontally. The side-length $2R$ of the square is equal to the diameter of each semi-circle, so that the area is determined by just one parameter, $R$. The cross-section is $2R$ (horizontal stadium) or $4R$ (vertical stadium).

\item a square, with rounded corners. The radius of curvature of the corners is set to one-eighth of the side-length of the square, $R=L/8$ , so that the area is again determined by just one parameter, and $H=L$. Also a diamond, which is the square rotated by $\pi/4$, with $H\approx \sqrt{2}L$.

\item a symmetric aerofoil, with long axis parallel to the direction of foam flow, defined by two arcs of circles bounded by two tangential straight lines. Three parameters are needed: length $L$ (distance between the centres of the circles), and radii $R_1$ (leading edge) and $R_2$ (trailing edge). This shape has up-down symmetry but not fore-aft symmetry, and cross-section $H=2 \max(R_1, R_2)$. If $R_1=R_2$, then this is a ``long'' horizontal stadium. 

\item an aerofoil-like shape with up-down asymmetry, in which two circles of equal radius $R_2$ are joined by arcs of radius $R_1$ and $R_1+2R_2$. The distance between the circles is parametrized by the angle $\theta_1$. Its cross-section is $H=(R_1+R_2)(1-\cos\theta_1)+2R_2$. This approximation to a standard aerofoil dispenses with the singular point at the trailing edge.

\end{enumerate}

We begin by describing our numerical method (\S \ref{sec:method}). The forces on each obstacle  are given in \S \ref{sec:resultsfo}; we find that the drag is mainly determined by its maximum cross-section $H$ perpendicular to the direction of flow and that a significant lift is found only for the aerofoil without up-down symmetry. The field of bubble pressure around the obstacle, which is the main contribution to this lift, is described in \S \ref{sec:resultsfi}, and we make some concluding remarks  in \S \ref{sec:concs}.

\section{Method}
\label{sec:method}

\begin{figure}
\centerline{
\subfigure[]{
\includegraphics[width=7cm,angle=0]{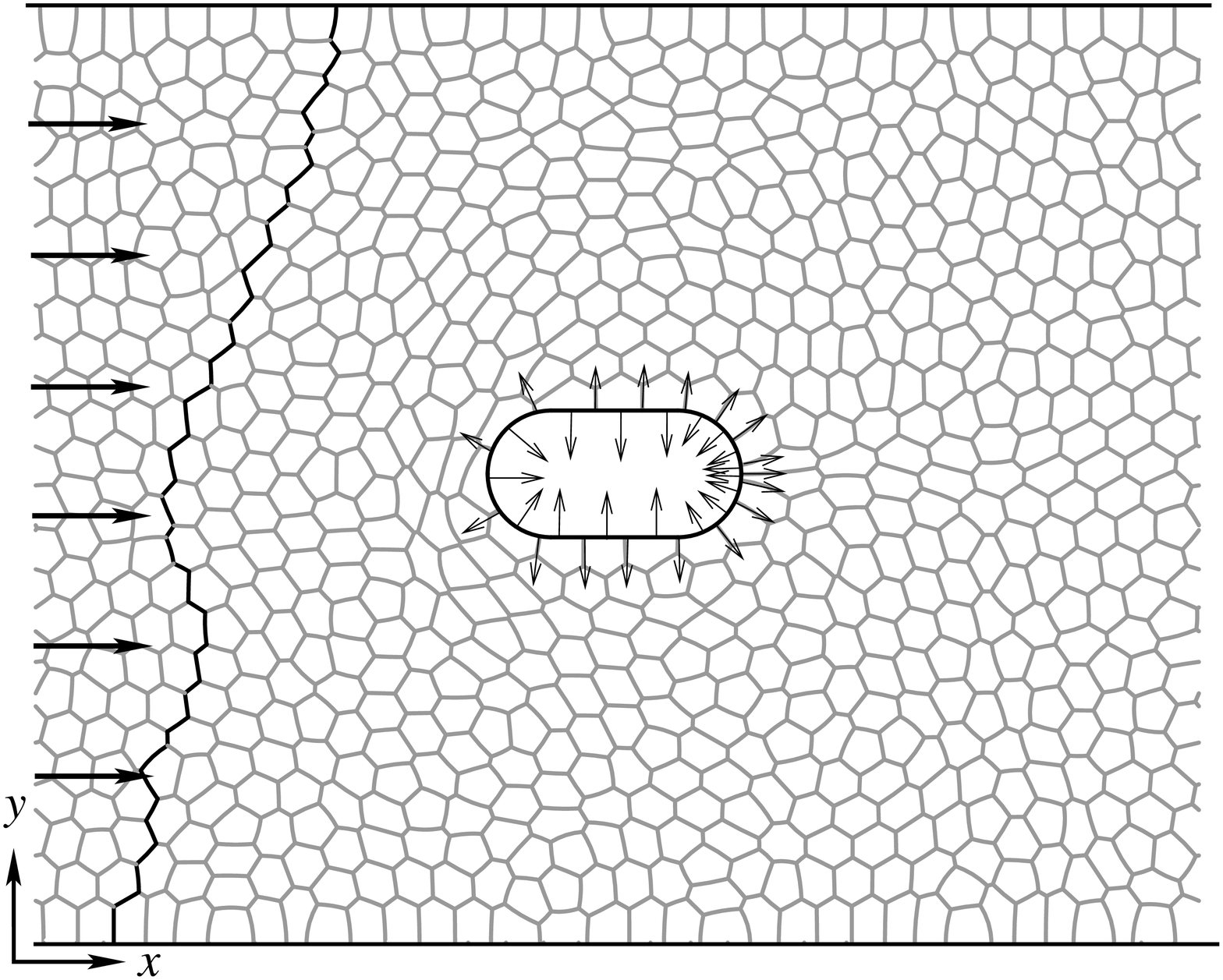}
}
\subfigure[]{
\includegraphics[width=8cm,angle=0]{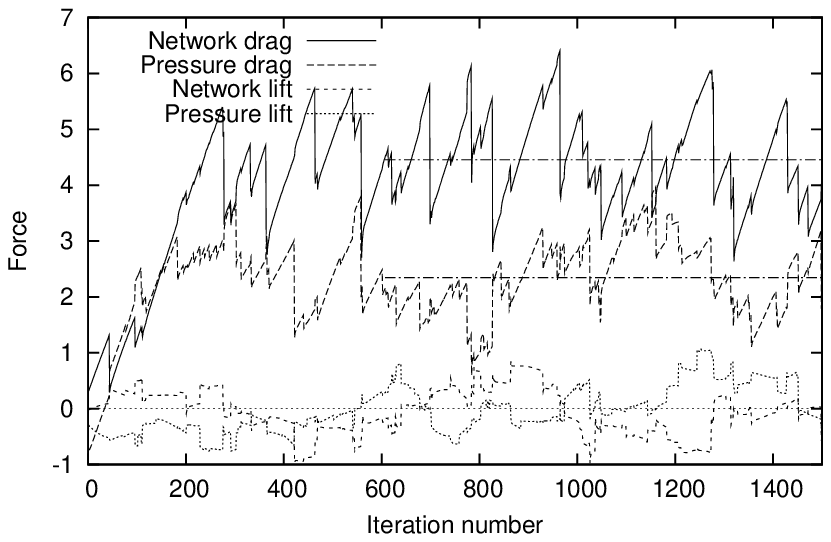}
}
}
\caption{(a) Sketch of the simulation, in this case for a horizontal stadium.  A 2D foam is created between two fixed walls and
caused to flow in the positive $x$ direction by increasing the area of the region to the left of the dark line of films joining the two walls. The obstacle is created in the centre of the channel; each film that touches the obstacle applies an equal
force outward in the direction normal to the obstacle and each bubble applies a pressure force inward at the middle of the
shared boundary. The films bunch up at the trailing edge of the obstacle and the bubble pressures rise at the leading edge due to the flow, leading to drag and lift forces on the obstacle. (b) Example (vertical stadium, area ratio $a_r = 6$) of the pressure ($F_P$) and network ($F_T$) contributions to the drag ($x$) and the lift ($y$) as a function of iteration number. The drag forces increase linearly before developing a saw-tooth variation which is linked to a build-up of stress followed by avalanches of T1s in the foam. The horizontal lines show the average drag forces. In this case the pressure and network contributions to the lift are both negligible. 
}
\label{fig:foam}
\end{figure}

We use the Surface Evolver \cite{brakke92} in the manner described by \citet{daviesc08}. 
We create three foams of around 725 bubbles (in this range the number of bubbles does not affect the results; data not shown) between parallel walls with a Voronoi construction \cite{brakke86,wyndc08}. The channel has unit length and width $W=0.8$. The foams are monodisperse, with bubble area denoted $A_b$ and about 22 bubbles in the cross-section of the channel. A bubble in the centre of the channel is chosen to represent the obstacle, and its periphery constrained to the required shape; its area is then increased until it reaches the desired area ratio $a_r = A_{obs}/A_{b}$ and it is then fixed -- see figure \ref{fig:foam}(a). The tension of each film, $\gamma$, which is twice the air-liquid surface tension and is in effect a line tension, is taken equal to one, without loss of generality.

The boundary conditions are that of free slip on the boundary of the obstacle and the channel walls, {\color{blue} so that the films meet the boundaries at $90^\circ$}, and periodicity in the direction of flow. We checked in a few instances that changing the boundary condition on the channel walls to non-slip has little effect on the forces on a small obstacle in the centre of the channel. At each iteration the foam is pushed with a small area increment $dA = 5\times 10^{-4}$ to create a pressure gradient \cite{raufastedcjg06}. The perimeter is then evolved towards a local minimum and T1s are performed whenever a film length shrinks below $l_c = 1 \times 10^{-3}$ (representing a foam with low liquid fraction, of the order of $10^{-4}$). A simulation runs for 1500 iterations to ensure that the measurements are made beyond any transient in which the foam retains a memory of its initial state. Each simulation takes about one week on a 1.5GHz CPU. The method has been validated against experiment in the case of an elliptical obstacle \cite{dolletdg06,daviesc10}.

\subsection{Drag and lift}

Each film that touches the obstacle applies an outward force with magnitude equal to the force of surface tension and direction perpendicular to the obstacle boundary. Their resultant is the network force
\begin{equation}
\vec{F}_T = \gamma \sum_{i} \vec{n}_i
\end{equation}
where $\vec{n}_i$ is the unit outward normal at the vertex $i$ terminating each film that meets the obstacle. See figure \ref{fig:foam}(a).

Each bubble that touches the obstacle applies a pressure force inward at the middle of the shared boundary. Their resultant is the pressure force
\begin{equation}
\vec{F}_P =  -\sum_{j}p_j l_j \vec{n}_j
\end{equation}
where $p_j$ is the pressure of bubble $j$, $l_j$ the length of  shared boundary and $\vec{n}_j$ the unit outward normal to the obstacle at the midpoint of the line joining the two ends of the shared boundary.

The drag on an obstacle is the component of the sum of the network and pressure forces in the direction of motion, $F_D = F_T^x + F_P^x$. The lift is the component perpendicular to this, $F_L = F_T^y + F_P^y$, with the convention that positive values of lift act in the positive $y$ direction. All four components are recorded at the end of each iteration, and averaged above 600 iterations, well beyond any transient. An example is shown in figure \ref{fig:foam}(b). The standard deviation of the fluctuations in force about this average are used to give the error bars in the figures below.
%The sensitivity to the starting value of the fit is around 1 or 2\% of the final average. 
%The rms error for each fit was found to be small -- of the order of the point size in the figures below -- so that we do not show error bars.

\begin{figure}

\end{figure}

\section{Results}

\subsection{Drag and lift force on an obstacle}
\label{sec:resultsfo}

The drag and lift oscillate in a saw-tooth fashion (figure \ref{fig:foam}(b)), caused by intervals in which the imposed strain is stored elastically followed by cascades of T1 topological changes. Nonetheless, they have a well-defined average. We find that for all obstacles with up-down symmetry the average lift is close to zero.

\begin{figure}
\centerline{
\subfigure[]{
\includegraphics[width=8cm,angle=0]{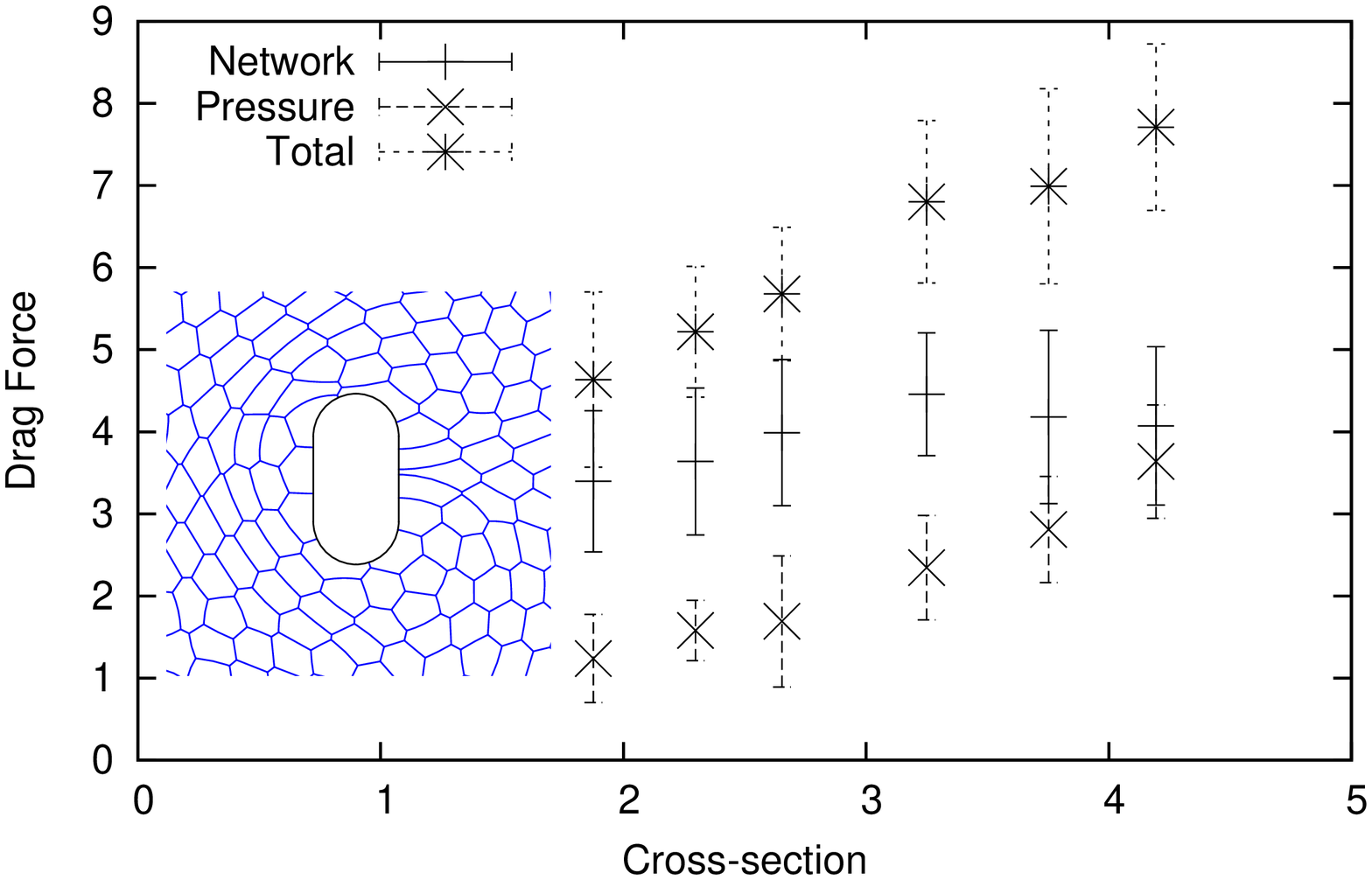}
}
\subfigure[]{
\includegraphics[width=8cm,angle=0]{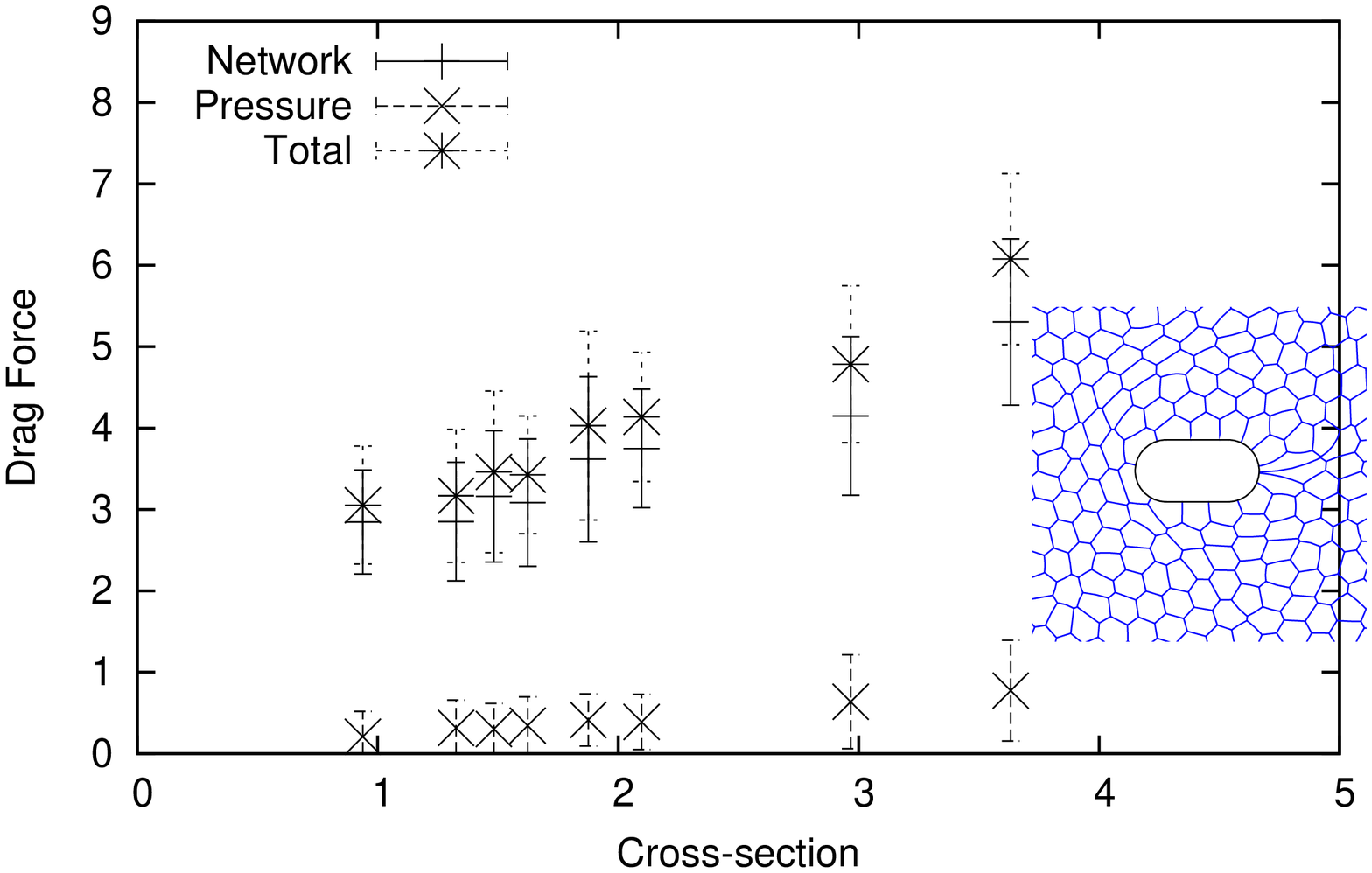}
}
}
\centerline{
\subfigure[]{
\includegraphics[width=8cm,angle=0]{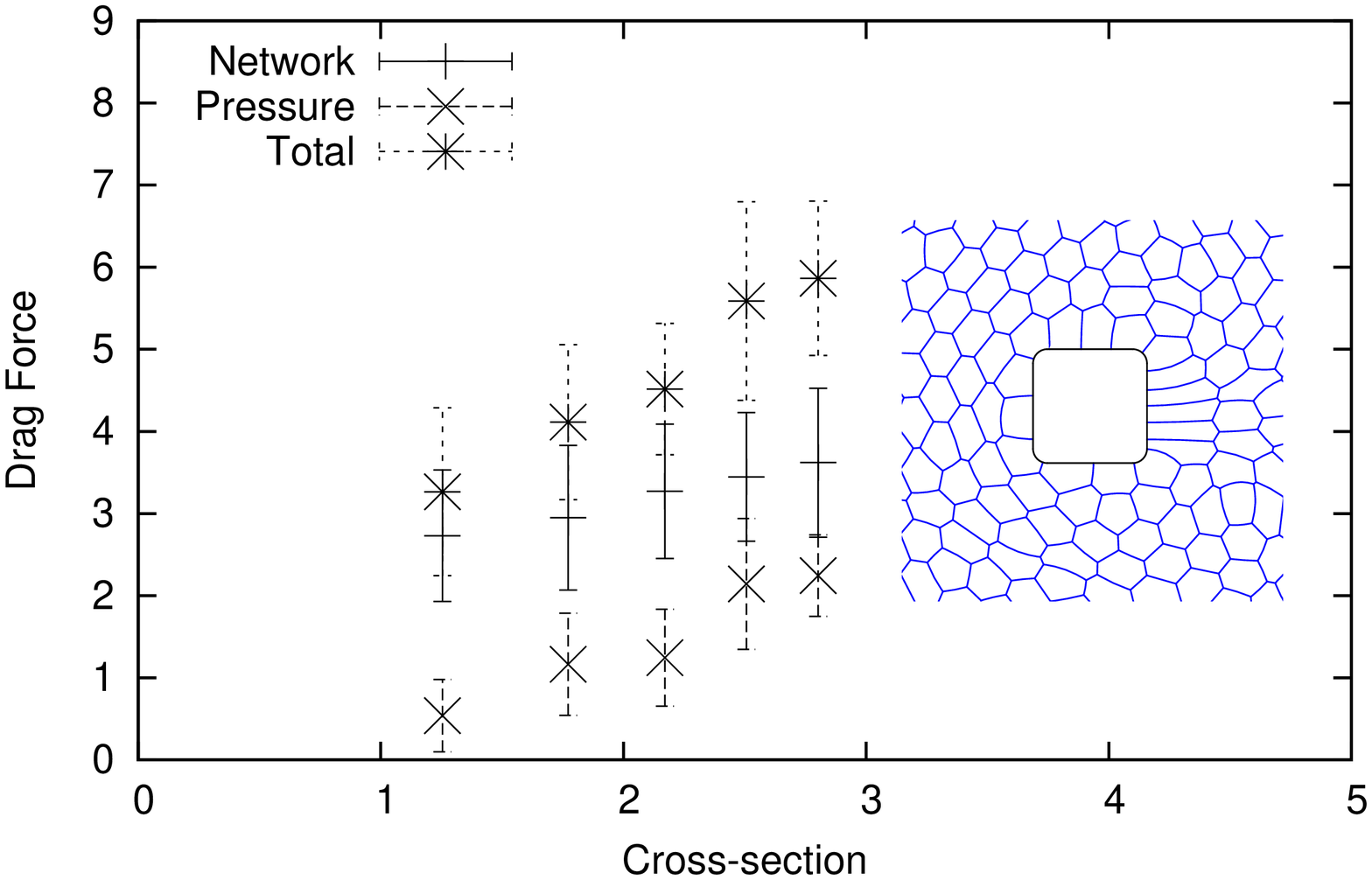}
}
\subfigure[]{
\includegraphics[width=8cm,angle=0]{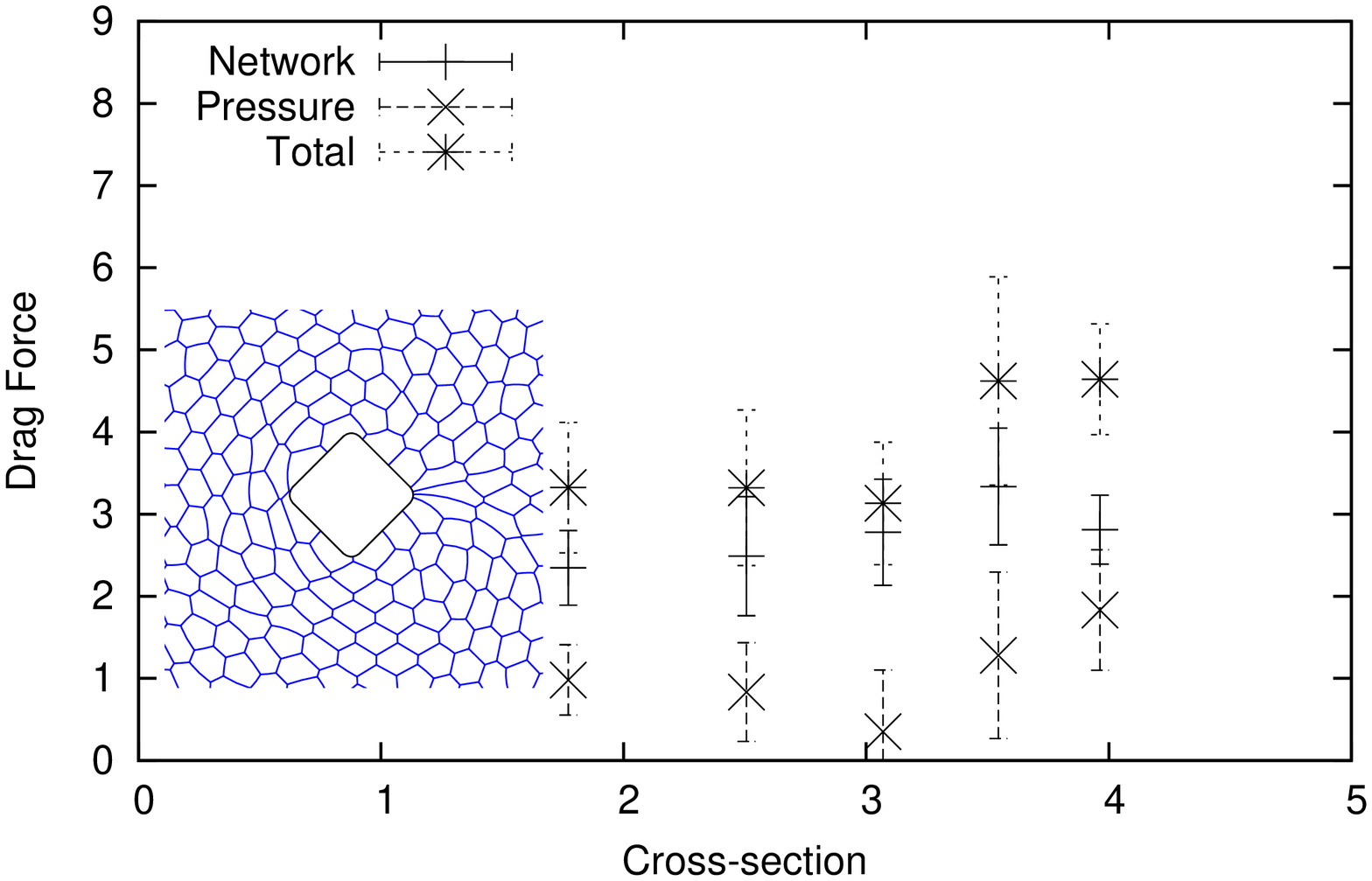}
}
}
\caption{(Color online) Drag {\em vs} obstacle cross-section $H/d_b$. Images are for obstacles with area ratio $a_r = 10$ with flow from left to right. 
(a) Vertical stadium ($a_r = 2,3,4,6,8,10$). 
(b) Horizontal stadium ($a_r = 2,4,5,6,8,10,20,30$). 
(c) Square ($a_r =2,4,6,8,10, L/R=8$). 
(d) Diamond ($a_r =2,4,6,8,10, L/R=8$). 
}
\label{fig:drag}
\end{figure}

We vary the area ratio of each obstacle, usually in the range one to ten but occasionally higher. 
We normalize the cross-section and length of each obstacle by the average bubble diameter $d_b = \sqrt{4A_b/\pi}$ which, since the walls are far enough away not to have an effect on the drag and lift, is the significant length-scale here. We choose to plot the resulting drag as a function of cross-section $H/d_b$ (figure \ref{fig:drag}) since it gives an approximately linear relationship \cite{raufastedcjg06}.
%Why shuold it be linear? Looks fairly linear versus area ratio
It is apparent that the drag increases with obstacle cross-section most quickly for ``blunt'' objects with a vertical leading edge (square, vertical stadium). Obstacles with a rounded leading edge (circle \cite{coxdg05}, horizontal stadium) experience lower drag for given cross-section. In each case, the main contribution to the drag is usually due to network forces; the pressure contribution to the total drag is lower but follows the same trends.

\begin{figure}
\centerline{
\subfigure[]{
\includegraphics[width=8cm,angle=0]{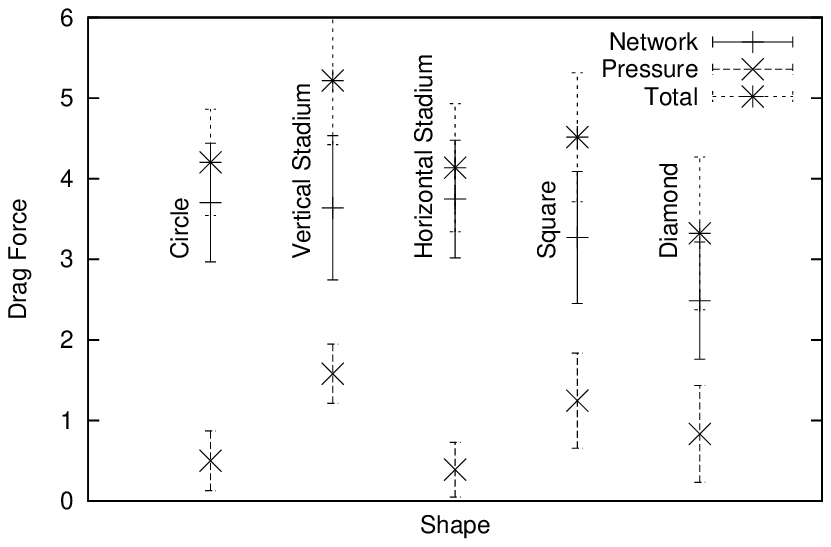}
}
\subfigure[]{
\includegraphics[width=8cm,angle=0]{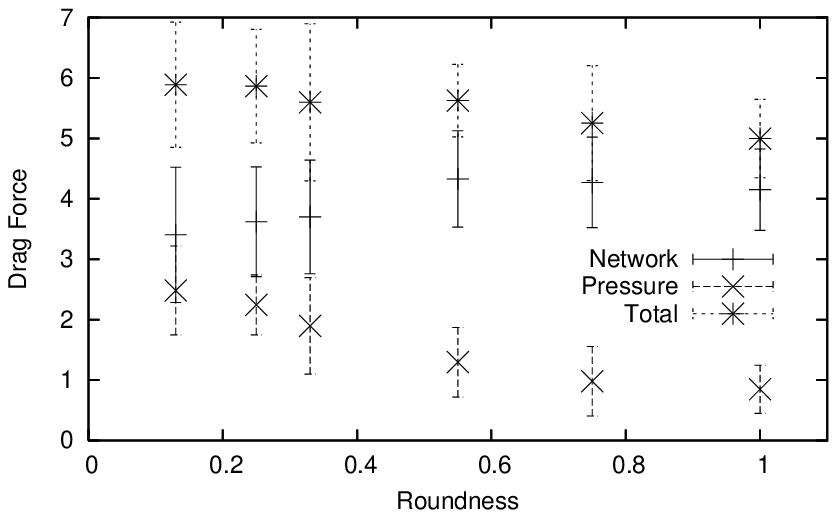}
}
}
\centerline{
\subfigure[]{
\includegraphics[width=8cm,angle=0]{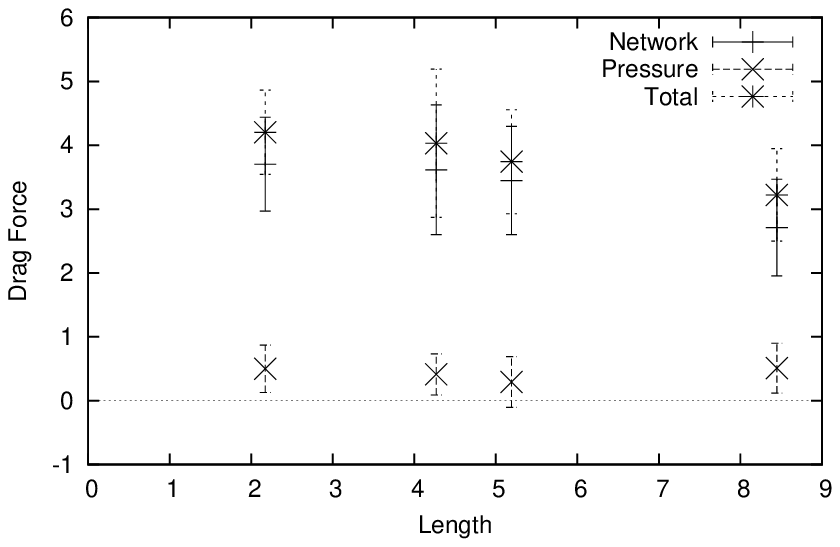}
}
\subfigure[]{
\includegraphics[width=8cm,angle=0]{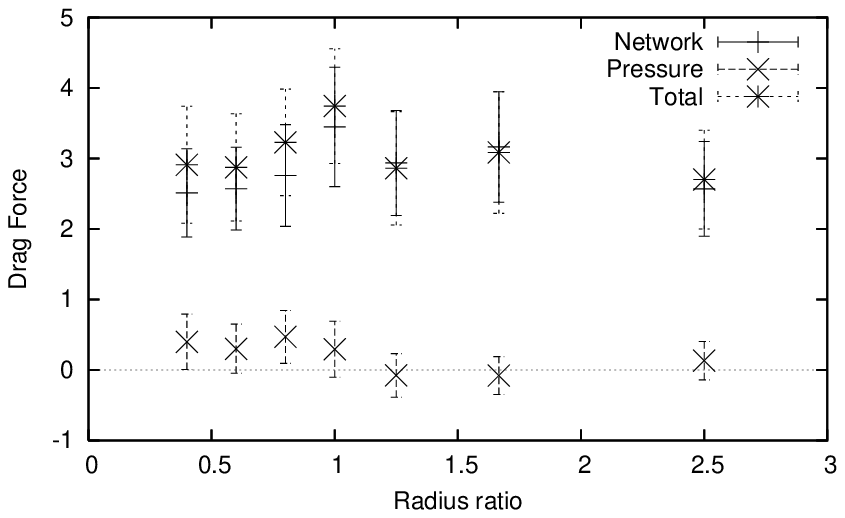}
}
}
\caption{Drag force on different obstacles. (a) Drag {\em vs} shape at constant cross-section $H/d_b \approx 2.1$. The pressure contribution to the drag decreases with the rounding of the leading edge and the network contribution decreases with the rounding of the trailing edge. (b) Drag {\em vs} roundness $R/(L+R)$, interpolating between a square ($R=L/8$) and a circle ($L=0$) with $a_r = 10$. The same effect is seen as in (a). (c) Drag {\em vs} obstacle length, measured as $(L+2R)/d_b$, for symmetric aerofoils with $R_1=R_2$ at constant cross-section $H/d_b \approx 2.1$. The first point on the left corresponds to a circle ($L=0$), and the second to a horizontal stadium ($L=2R$).  The network contribution to the drag decreases slightly with length. (d) Drag {\em vs} radius ratio $R_2/R_1$ for a symmetric aerofoil ($a_r = 10, L$ varies). The pressure drag decreases when the leading edge has a smaller radius of curvature.   
}
\label{fig:drag-length}
\end{figure}
%A sharp leading edge/ rounded trailing edge reduces the pressure drag and increases the network drag.

To tease out the effect of obstacle shape on the two components of drag studied here, we fix the cross-section (figure \ref{fig:drag-length}(a),(b)) and vary the shape. The pressure contribution to the drag is highest when the leading edge is blunt (vertical stadium, square), since this causes the greatest deformation to the bubbles. Similarly, the network contribution to the drag is highest when the trailing edge is rounded (the most ``circular'' case in figure \ref{fig:drag-length}(b)), although this effect is weaker, since a rounded trailing edge allows more films to collect in that area. The shape of the diamond is such that the network drag is very low, since films can gather on the sloping sides as well as the rounded region at the very tip of the trailing edge, while the pressure drag is intermediate.

 The length $L$ of an obstacle has only a weak effect on the drag (figure \ref{fig:drag-length}(c)). In particular, this is the case for a symmetric aerofoil with $R_2=R_1$, since most of the films that touch the obstacle are perpendicular to the direction of foam flow. By varying the ratio $R_2/R_1$ for a symmetric aerofoil with fixed cross-section $H$ and fixed area ratio $a_r = 10$, we can investigate the effect of fore-aft asymmetry. Figure \ref{fig:drag-length}(d) shows that the total drag varies little, emphasizing that cross-section and rounded leading and trailing edges make the major contribution to the drag. The pressure contribution to the drag decreases with $R_2/R_1$, that is, as the leading edge gets smaller and bubbles are less deformed there.

\begin{figure}
\centerline{
\includegraphics[width=8cm,angle=0]{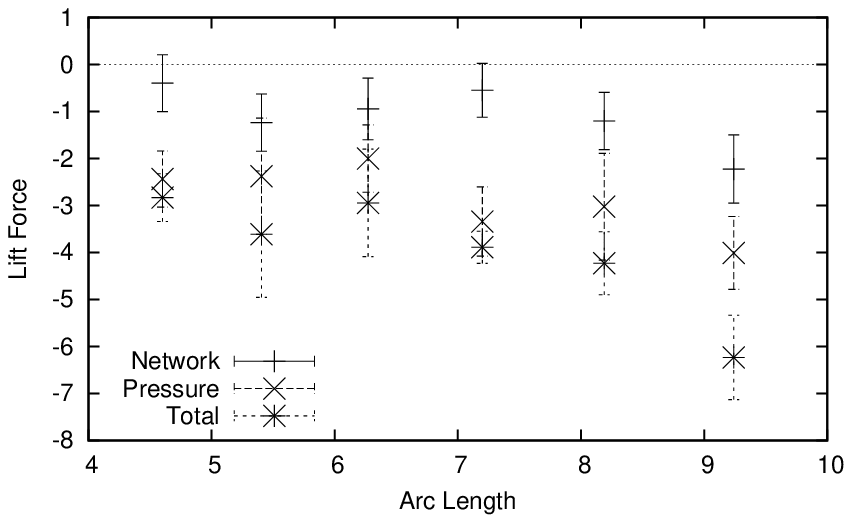}
}
\caption{Lift versus asymmetric aerofoil arc length, scaled by $d_b$. All three of $R_1$, $R_2$ and $\theta_1$ are chosen to increase roughly in the same proportion. The lift is always in the negative $y$ direction and the network contribution  is smaller than that due to pressure.} 
\label{fig:liftaerofoil}
\end{figure}

The lift is, on average, zero for all obstacles with a horizontal axis of symmetry (as in figure \ref{fig:foam}(b)); it is only significant for the asymmetric aerofoil, being negative and of the same order of magnitude as the drag. In particular the lift increases with aerofoil length (figure \ref{fig:liftaerofoil}), and the major component of lift arises from the bubble pressures. It appears therefore that the curvature of the aerofoil induces changes in bubble pressures, and that it is this, rather than an imbalance in the number of films pulling on the top and bottom surfaces of the object, that gives rise to the lift. We return to the bubble pressures below.

To test the effect of obstacle position in the channel, we placed the same asymmetric aerofoil in three different positions across the channel: $y = 0.25W,  0.5W$ (reference case) and $0.75W$. No significant difference in the drag or lift was observed (data not shown), indicating that the obstacle was still sufficiently far from the walls that they don't interfere with the flow (recall that this is a elasto-plastic rather than a viscous flow, distinct from a Newtonian fluid where the wall always has an effect in 2D) and that the lift is not just due to the foam squeezing through the gap between wall and obstacle.

\begin{figure}
\centerline{
\subfigure[]{
\includegraphics[width=0.45\textwidth,angle=270]{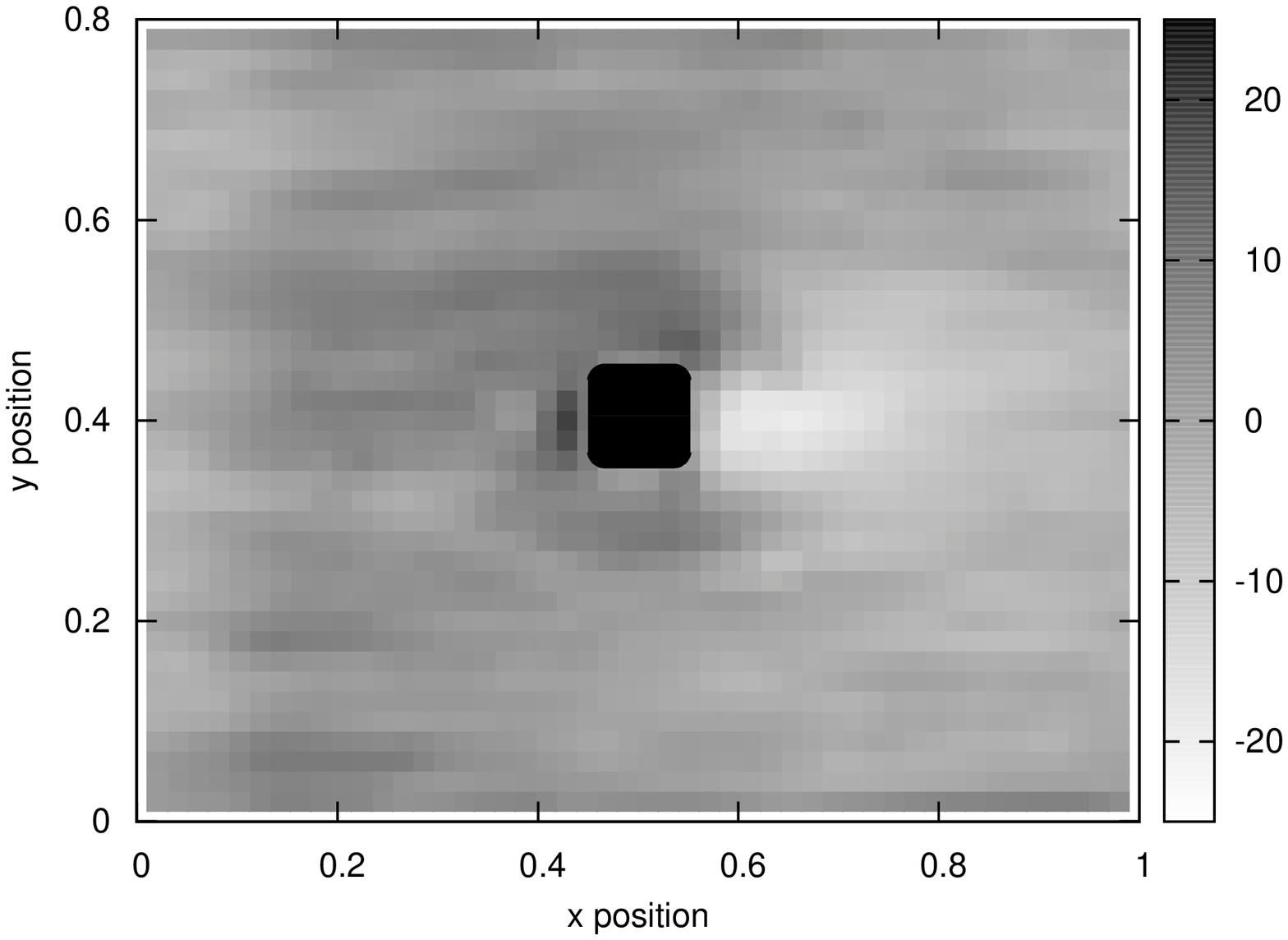}
}
\subfigure[]{
\includegraphics[width=0.45\textwidth,angle=270]{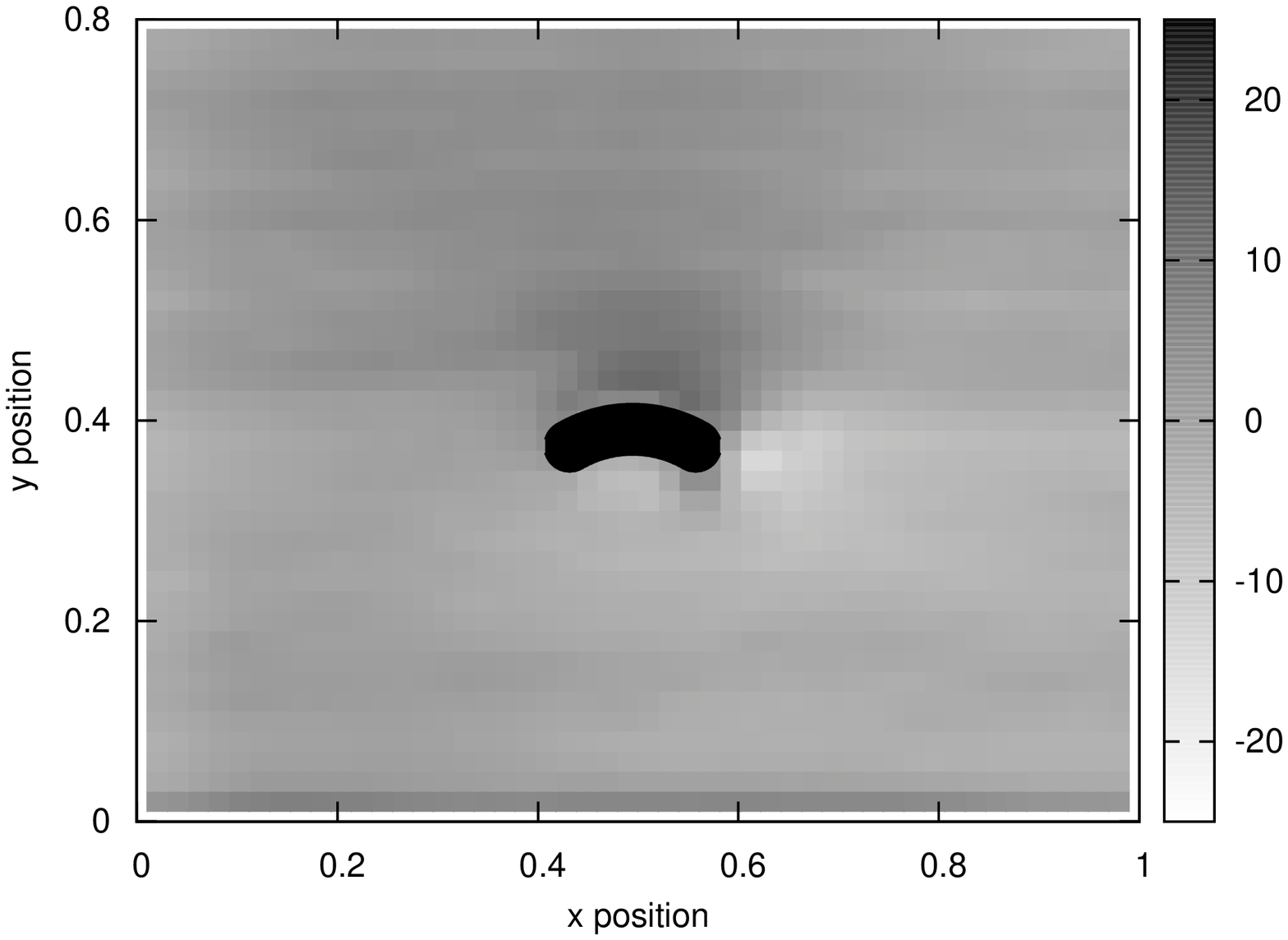}
}
}
\centerline{
\subfigure[]{
\includegraphics[width=7cm,angle=0]{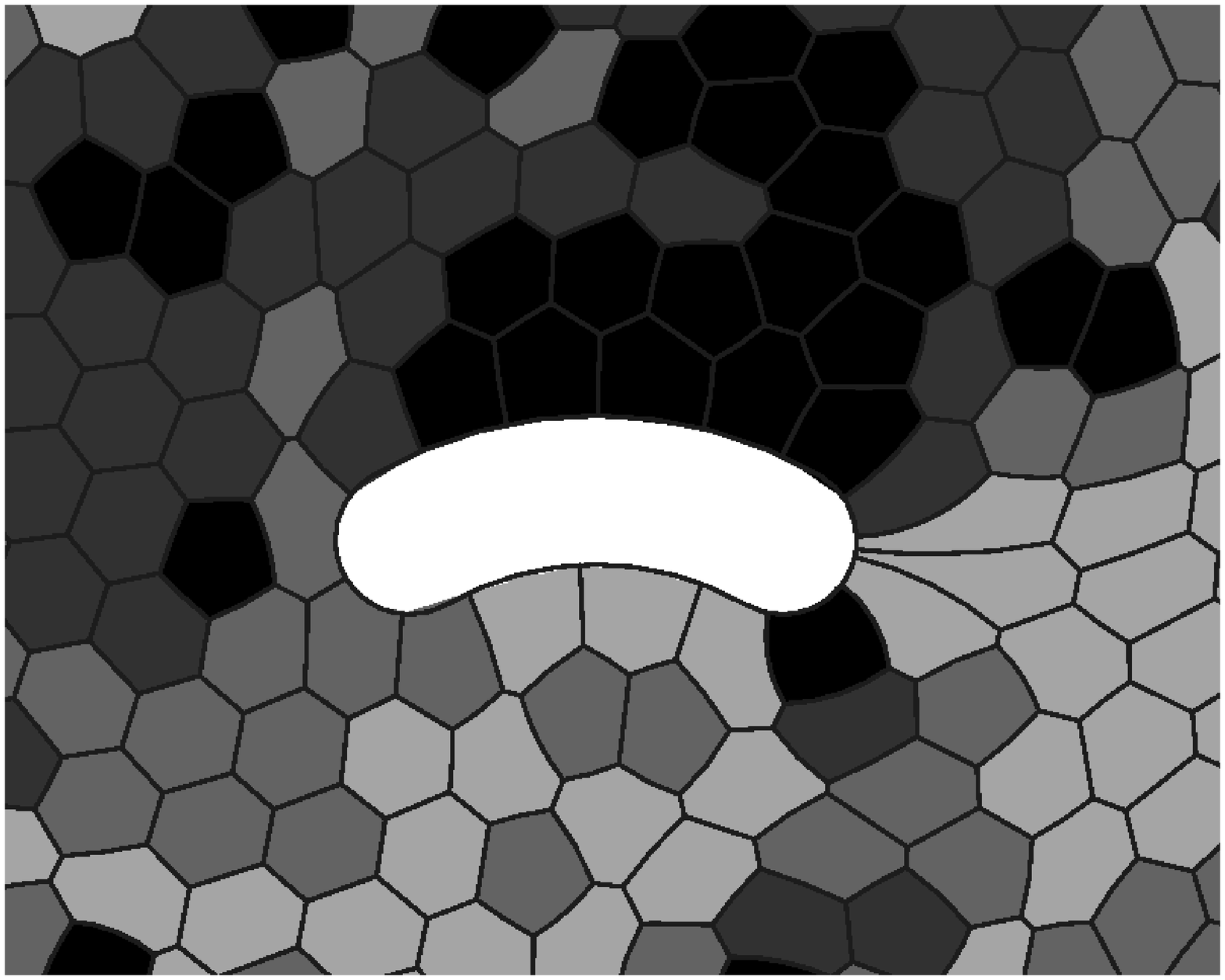}
}
}
\caption{Pressure fields averaged over the duration of the simulation. (a) Square obstacle with $a_r = 10$, showing increased pressure upstream of the obstacle and low pressure downstream. (b) Asymmetric aerofoil, with $R_1/\sqrt{d_b}= 3$, $R_2/\sqrt{d_b} = 0.75$ and $\theta_1 =\pi/6$, showing low pressure beneath as well as downstream. The increase of pressure upstream is less-pronounced, and there is a pressure peak beneath the trailing edge of the aerofoil. (c) Zoom of the typical arrangement of films around the same aerofoil, with bubbles shaded by instantaneous pressure on a scale by which pressure increases with grey intensity. In both representations a region of low pressure is evident beneath the aerofoil -- it is this which induces a negative lift -- as well as the pressure peak beneath the trailing edge.}
\label{fig:fields_p}
\end{figure}

\subsection{Pressure field around an obstacle}
\label{sec:resultsfi}

To further probe the phenomenon of negative lift in  foams, in figure \ref{fig:fields_p} we compare the distribution of bubble pressures around the flat-bottomed aerofoil with an up-down symmetric obstacle typified by the square. The Surface Evolver calculates the bubble pressures (as Lagrange multipliers of the area constraints) in such a way that they are all relative to the pressure of one bubble. Thus the average pressure is subtracted from all values at each iteration, before binning the data as above.

The bubble pressures decrease in the $x$ direction, on average, because of the flow. The presence of an obstacle induces a region of high pressure at the leading edge and a region of low pressure at the trailing edge. In addition, the asymmetric aerofoil shows a region of high pressure above and low pressure below, confirming that the pressure contribution to the lift is downwards.

\section{Conclusions}
\label{sec:concs}

\begin{figure}
\centerline{
\subfigure[]{
\includegraphics[width=8cm,angle=0]{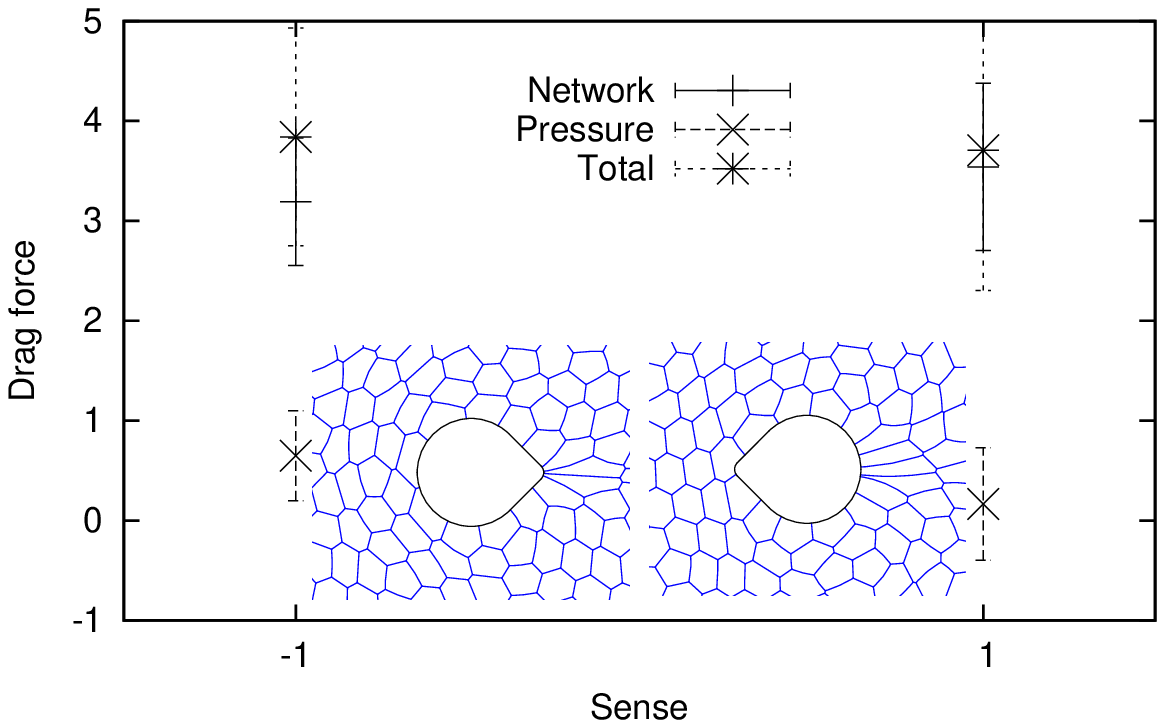}
}
\subfigure[]{
\includegraphics[width=7cm,angle=0]{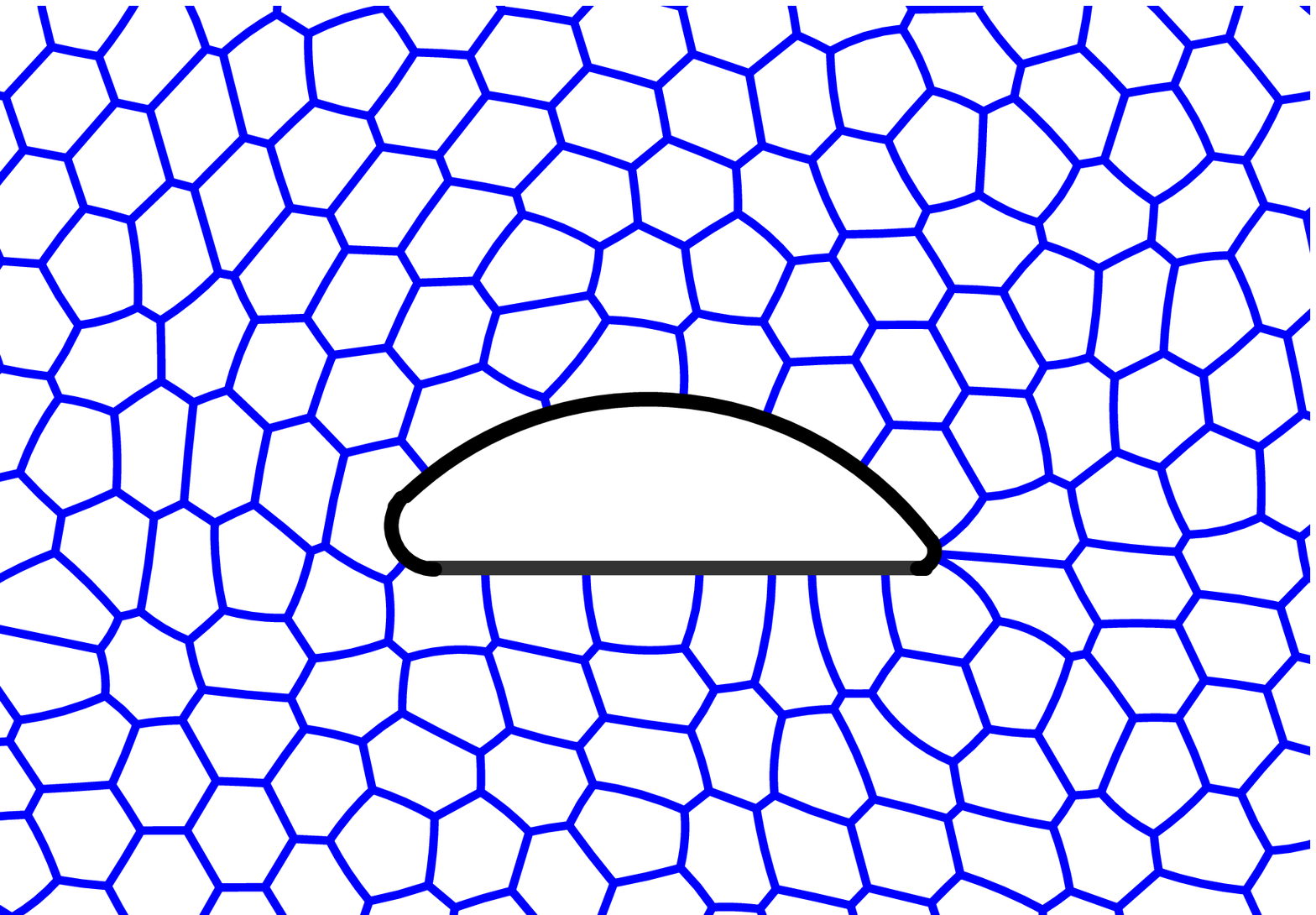}
}
}
\caption{(Color online) (a) An interpolation between a diamond and a circle takes the shape shown, with $a_r = 10$. When its sense is flipped relative to the direction of flow, the relative contributions to the pressure and network drag change, while the total drag remains the same: a sharp leading edge and rounded trailing edge reduces the pressure drag and increases the network drag. (b) Instantaneous arrangement of films around a flat-bottomed aerofoil (parameters: $a_r = 9$, cross-section $H/d_b = 1.59$, radius of curvature of leading edge is $R_1/d_b = 0.41$, of trailing edge is $R_2/d_b = 0.17$ and of upper side is $R_3/d_b=3.46$). The instantaneous values of drag and lift are $F_y^T = -2.90, F_y^P = -2.36, F_x^T = 2.03, F_x^P = 0.42$. The lift is again negative, both network and pressure contributions are similar, and the total lift is of the same order of magnitude as the total drag.
}
\label{fig:futureshapes}
\end{figure}

The simulations described here show that the forces on an obstacle embedded in a flow of foam depend strongly on the shape of the obstacle. We separate two components, due to the pressure in the bubbles and the network of soap films, and find that the pressure contribution decreases with the rounding of the leading edge and the network contribution decreases with the rounding of the trailing edge. Further evidence is given in figure \ref{fig:futureshapes}(a).

%R1 = 0.132;R2 = 0.0158;R3 = 0.00632;area ratio = 9.0755;length = 0.217;Thickness = 0.0608;

In classical fluid mechanics, the presence of viscosity can give rise to trailing vortices and circulation around an obstacle in a fluid flow. Here, not only do we neglect viscosity, but the discrete nature of the foam probably suppresses any possibility of circulation. Yet a lift force is still observed for obstacles without lateral symmetry, and it arises because of the way in which the obstacle deforms the bubbles that make up the foam. It is therefore an effect of elasticity or, more generally, viscoelasticity \cite{wangj04,wangjosephairfoil}, due to the normal stresses generated in the fluid, and acts in the opposite direction to the usual sense of ``lift''. A concave underside, as in the familiar Joukowski profile and the asymmetric aerofoil described above, is not necessary to obtain a negative lift (figure \ref{fig:futureshapes}(b)).

It remains to determine whether a given obstacle is actually stable with respect to rotation; that is, whether the torque on any given obstacle is sufficient to rotate it and thereby reduce the drag and/or lift. This is a necessary pre-cursor to using this work to determining which shapes of obstacles offer the least resistance to foam flow. It is also of interest to incorporate some element of viscous dissipation, perhaps using the viscous froth model \cite{kernwahc03}, within the simulations, which has a particularly significant effect on rotation \cite{daviesc10} but also the film motion around an obstacle. We shall return to both these issues in future work.

\begin{acknowledgments}
We thank K. Brakke for developing, distributing and supporting the Surface Evolver, I.T. Davies for technical assistance with the simulations, and F. Graner for useful comments. SJC acknowledges financial support from EPSRC (EP/D071127/1).
\end{acknowledgments}

%\bibliography{/home/sxc/LATEX/BIB/foam,/home/sxc/LATEX/BIB/weaire,/home/sxc/LATEX/BIB/tmp}

\begin{thebibliography}{28}
\providecommand{\natexlab}[1]{#1}
\providecommand{\url}[1]{\texttt{#1}}
\expandafter\ifx\csname urlstyle\endcsname\relax
  \providecommand{\doi}[1]{doi: #1}\else
  \providecommand{\doi}{doi: \begingroup \urlstyle{rm}\Url}\fi

\bibitem[Prud'homme and Khan(1996)]{prudk96}
R.K. Prud'homme and S.A. Khan, editors.
\newblock \emph{{Foams: Theory, Measurements and Applications}}, volume~{\bf
  57} of \emph{{Surfactant Science Series}}.
\newblock Marcel Dekker, New York, 1996.

\bibitem[Weaire and Hutzler(1999)]{WeaireH99}
D.~Weaire and S.~Hutzler.
\newblock \emph{{The Physics of Foams}}.
\newblock {Clarendon Press, Oxford}, 1999.

\bibitem[H{\"o}hler and Cohen-Addad(2005)]{hohlerca05}
R.~H{\"o}hler and S.~Cohen-Addad.
\newblock {Rheology of Liquid Foam}.
\newblock \emph{{J. Phys.: Condens. Matter}}, {\bf{17}}:\penalty0 R1041--R1069,
  2005.

\bibitem[Cantat et~al.(2010)Cantat, Cohen-Addad, Elias, Graner, H{\"o}hler,
  Pitois, Rouyer, and Saint-Jalmes]{mousse10}
I.~Cantat, S.~Cohen-Addad, F.~Elias, F.~Graner, R.~H{\"o}hler, O.~Pitois,
  F.~Rouyer, and A.~Saint-Jalmes.
\newblock \emph{{Les mousses - structure et dynamique}}.
\newblock Belin, Paris, 2010.

\bibitem[Stokes(1850)]{stokes50}
G.G. Stokes.
\newblock {On the Effect of the Internal Friction of Fluids on the Motion of
  Pendulums}.
\newblock \emph{{Trans. Camb. Phil. Soc.}}, {\bf{IX}}:\penalty0 8--149, 1850.

\bibitem[Cox et~al.(2000)Cox, Alonso, Hutzler, and Weaire]{coxahw00}
S.J. Cox, M.D. Alonso, S.~Hutzler, and D.~Weaire.
\newblock The Stokes experiment in a foam.
\newblock In {P. Zitha, J. Banhart and G. Verbist}, editor, \emph{Foams,
  emulsions and their applications}, pages 282--289. MIT-Verlag, Bremen, 2000.

\bibitem[Asipauskas et~al.(2003)Asipauskas, Aubouy, Glazier, Graner, and
  Jiang]{asipausaggj03}
M.~Asipauskas, M.~Aubouy, J.A. Glazier, F.~Graner, and Y.~Jiang.
\newblock {A texture tensor to quantify deformations: the example of
  two-dimensional flowing foams}.
\newblock \emph{Granular Matter}, {\bf{5}}:\penalty0 71--74, 2003.

\bibitem[{de Bruyn}(2004)]{bruyn04}
J.R. {de Bruyn}.
\newblock {Transient and steady-state drag in foam}.
\newblock \emph{{Rheol. Acta}}, {\bf{44}}:\penalty0 150--159, 2004.

\bibitem[Dollet et~al.(2005{\natexlab{a}})Dollet, Aubouy, and
  Graner]{dolletag04}
B.~Dollet, M.~Aubouy, and F.~Graner.
\newblock {Inverse Lift: a signature of the elasticity of complex fluids}.
\newblock \emph{{Phys. Rev. Lett.}}, {\bf{95}}:\penalty0 168303,
  2005{\natexlab{a}}.

\bibitem[Dollet et~al.(2005{\natexlab{b}})Dollet, Elias, Quilliet, Raufaste,
  Aubouy, and Graner]{dolleteqrag05}
B.~Dollet, F.~Elias, C.~Quilliet, C.~Raufaste, M.~Aubouy, and F.~Graner.
\newblock {Two-dimensional flow of foam around an obstacle: Force
  measurements}.
\newblock \emph{{Phys. Rev. E}}, {\bf{71}}:\penalty0 031403,
  2005{\natexlab{b}}.

\bibitem[Cantat and Pitois(2005)]{cantatp05}
I.~Cantat and O.~Pitois.
\newblock {Mechanical probing of liquid foam ageing}.
\newblock \emph{{J. Phys.: Condens. Matter}}, {\bf{17}}:\penalty0 S3455--S3461,
  2005.

\bibitem[Dollet et~al.(2005{\natexlab{c}})Dollet, Elias, Quilliet, Huillier,
  Aubouy, and Graner]{dolleteqhag05}
B.~Dollet, F.~Elias, C.~Quilliet, A.~Huillier, M.~Aubouy, and F.~Graner.
\newblock {Two-dimensional flows of foam: drag exerted on circular obstacles
  and dissipation}.
\newblock \emph{{Coll. Surf. A}}, {\bf{263}}:\penalty0 101--110,
  2005{\natexlab{c}}.

\bibitem[Cox et~al.(2006)Cox, Dollet, and Graner]{coxdg05}
S.J. Cox, B.~Dollet, and F.~Graner.
\newblock {Foam flow around an obstacle: simulations of obstacle-wall
  interaction}.
\newblock \emph{{Rheol. Acta.}}, {\bf{45}}:\penalty0 403--410, 2006.

\bibitem[Cantat and Pitois(2006)]{cantatp06}
I.~Cantat and O.~Pitois.
\newblock {Stokes experiment in a liquid foam}.
\newblock \emph{{Phys. Fluids}}, {\bf{18}}:\penalty0 083302, 2006.

\bibitem[Dollet et~al.(2006)Dollet, Durth, and Graner]{dolletdg06}
B.~Dollet, M.~Durth, and F.~Graner.
\newblock {Flow of foam past an elliptical obstacle}.
\newblock \emph{{Phys. Rev. E}}, {\bf{73}}:\penalty0 061404, 2006.

\bibitem[Dollet and Graner(2007)]{dolletg06}
B.~Dollet and F.~Graner.
\newblock {Two-dimensional flow of foam around a circular obstacle: local
  measurements of elasticity, plasticity and flow}.
\newblock \emph{{J. Fl. Mech.}}, {\bf{585}}:\penalty0 181--211, 2007.

\bibitem[Raufaste et~al.(2007)Raufaste, Dollet, Cox, Jiang, and
  Graner]{raufastedcjg06}
C.~Raufaste, B.~Dollet, S.~Cox, Y.~Jiang, and F.~Graner.
\newblock {Yield drag in a two-dimensional foam flow around a circular
  obstacle: Effect of liquid fraction}.
\newblock \emph{{Euro. Phys. J. E}}, {\bf{23}}:\penalty0 217--228, 2007.

\bibitem[Tabuteau et~al.(2007)Tabuteau, Oppong, {de Bruyn}, and
  Coussot]{tabuteauobc07}
H.~Tabuteau, F.K. Oppong, J.R. {de Bruyn}, and P.~Coussot.
\newblock Drag on a sphere moving through an aging system.
\newblock \emph{{Europhys. Lett.}}, {\bf{78}}:\penalty0 68007, 2007.

\bibitem[Wyn et~al.(2008)Wyn, Davies, and Cox]{wyndc08}
A.~Wyn, I.T. Davies, and S.J. Cox.
\newblock Simulations of two-dimensional foam rheology: localization in linear
  couette flow and the interaction of settling discs.
\newblock \emph{{Euro. Phys. J. E}}, {\bf{26}}:\penalty0 81--89, 2008.

\bibitem[Davies and Cox(2009)]{daviesc08}
I.T. Davies and S.J. Cox.
\newblock Sedimenting discs in a two-dimensional foam.
\newblock \emph{{Coll. Surf. A}}, {\bf{344}}:\penalty0 8--14, 2009.

\bibitem[Bragg and Nye({1947})]{braggn47}
L.~Bragg and J.F. Nye.
\newblock {A dynamical model of a crystal structure}.
\newblock \emph{{Proc. R. Soc. Lond.}}, {A\bf{190}}:\penalty0 474--481, {1947}.

\bibitem[Weaire and Rivier(1984)]{weairer84}
D.~Weaire and N.~Rivier.
\newblock Soap, cells and statistics -- random patterns in two dimensions.
\newblock \emph{{Contemp. Phys.}}, {\bf{25}}:\penalty0 59--99, 1984.

\bibitem[Buzza et~al.(1995)Buzza, Lu, and Cates]{buzzlc95}
D.M.A. Buzza, C.-Y.~D. Lu, and M.E. Cates.
\newblock {Linear Shear Rheology of Incompressible Foams}.
\newblock \emph{{J. Phys. II France}}, {\bf{5}}:\penalty0 37--52, 1995.

\bibitem[Davies and Cox(2010)]{daviesc10}
I.T. Davies and S.J. Cox.
\newblock Sedimentation of an elliptical object in a two-dimensional foam.
\newblock \emph{{J. Non-Newt. Fl. Mech.}}, {\bf{165}}:\penalty0 793--799, 2010.

\bibitem[Brakke(1992)]{brakke92}
K.~Brakke.
\newblock {The Surface Evolver}.
\newblock \emph{{Exp. Math.}}, {\bf{1}}:\penalty0 141--165, 1992.

\bibitem[Brakke(1986)]{brakke86}
K.~Brakke.
\newblock {200,000,000 Random Voronoi Polygons}.
\newblock www.susqu.edu/brakke/papers/voronoi.htm, 1986.
\newblock {Unpublished}.

\bibitem[Wang and Joseph(2004)]{wangj04}
J.~Wang and D.D. Joseph.
\newblock Potential flow of a second-order fluid over a sphere or an ellipse.
\newblock \emph{{J. Fl. Mech.}}, {\bf{511}}:\penalty0 201--215, 2004.

\bibitem[Wang and Joseph(2005)]{wangjosephairfoil}
J.~Wang and D.D. Joseph.
\newblock The lift, drag and torque on an airfoil in foam modeled by the
  potential flow of a second-order fluid, 2005.
\newblock www.aem.umn.edu/people/faculty/joseph/archive/docs/931$\_$airfoilfoam.pdf.
\newblock Unpublished.


\bibitem[Kern et~al.(2004)Kern, Weaire, Martin, Hutzler, and Cox]{kernwahc03}
N.~Kern, D.~Weaire, A.~Martin, S.~Hutzler, and S.J. Cox.
\newblock {Two-dimensional viscous froth model for foam dynamics}.
\newblock \emph{{Phys. Rev. E}}, {\bf{70}}:\penalty0 041411, 2004.

\end{thebibliography}
%\bibliographystyle{unsrtnat}

\end{document}